\documentclass{optica-article}

\journal{opticajournal} % for journals or Optica Open

\articletype{Research Article}
\usepackage{lineno}
\usepackage{multicol}
\usepackage{amsmath}
\usepackage{graphicx}% Include figure files
\usepackage{dcolumn}% Align table columns on decimal point
\usepackage{xcolor}

\usepackage[T1]{fontenc}
\usepackage{ragged2e}
\usepackage{setspace}
\usepackage{times}
\usepackage{soul}
\usepackage{geometry}
\usepackage{hyperref} % To use the \url command (in the footnote)
\usepackage{marginnote}
\usepackage[utf8]{inputenc}
\usepackage{textcomp}

\usepackage{amsmath}
\usepackage{mathtools}
\usepackage{wrapfig}

\fancypagestyle{plain}{%
    \fancyhf{}%
    \fancyfoot[R]{\thepage}%
}

\begin{document}

\title{High-dimensional quantum correlation measurements with an adaptively gated hybrid single-photon camera}

\author{Sanjukta Kundu, Jerzy Szuniewicz, Grzegorz Firlik, Alexander Krupinski-Ptaszek and Radek Lapkiewicz\authormark{*}}

\address{Institute of Experimental Physics, Faculty of Physics, University of Warsaw, ul. Pasteura 5,02-093 Warszawa, Poland\\}

\email{\authormark{*}Corresponding author: radek.lapkiewicz@fuw.edu.pl}

\begin{abstract*} 
 
Efficient measurement of high-dimensional quantum correlations, especially spatial ones, is essential for quantum technologies, given their inherent high dimensionality and easy manipulation with basic optical elements. We propose and demonstrate an adaptively-gated hybrid intensified camera (HIC) that combines the information from a high spatial resolution sensor and a high temporal resolution detector, offering precise control over the number of photons detected within each frame. The HIC facilitates spatially resolved single-photon counting measurements. We study the measurement of momentum correlations of photon pairs generated in type-I spontaneous parametric down-conversion with the HIC and demonstrate the possibility of time-tagging the registered photons. With a spatial resolution of nearly 9 megapixels and nanosecond temporal resolution, this system allows for the realization of previously infeasible quantum optics experiments.

\end{abstract*}

%%%%%%%%%%%%%%%%%%%%%%%%%%  body  %%%%%%%%%%%%%%%%%%%%%%%%%%
\begin{multicols}{2}
\section{Introduction}
 High-dimensional quantum correlations are becoming extensively explored resources to facilitate quantum communication~\cite{Walborn2006, PhysRevA.77.062323, Walborn2007, PhysRevLett.123.070505}, sensing~\cite{PhysRevX.3.011013, Farooq2022} and computation~\cite{PhysRevA.83.052325}. Compared to qubits, high-dimensional systems offer larger information capacity~\cite{BechmannPasquinucci2000}, increased robustness against losses~\cite{Ecker2019}, and improved computational power~\cite{Reimer2018}. However, to harness their full potential, it is essential to measure and manipulate the high-dimensional states in different degrees of freedom~\cite{Erhard2020, Malik2016, Krenn2014}. To that end, the spatial degree of freedom of photonic systems has emerged as a promising avenue for exploring high-dimensional quantum correlations. Significant advancements have been made in the fields of quantum imaging~\cite{Defienne2021, Moreau2019}, metrology~\cite{Tse2019}, and the testing of fundamental principles of quantum mechanics~\cite{PhysRevLett.104.060401} through the measurement of spatial correlations between photon pairs generated by spontaneous parametric down-conversion (SPDC).

Characterizing photon pairs that are correlated in their transverse spatial modes requires highly sensitive detectors capable of resolving individual photons both in time and position which makes it a challenging task. Such measurements require high acquisition rates to collect statistically significant data in a short time span. Moreover, achieving accurate observations of real coincidences and distinguishing them from accidental coincidences require good temporal resolution.

\begin{figure*}[ht!]
\centering
    \captionsetup{width=\linewidth}
    \centering
    \includegraphics[width=\linewidth]{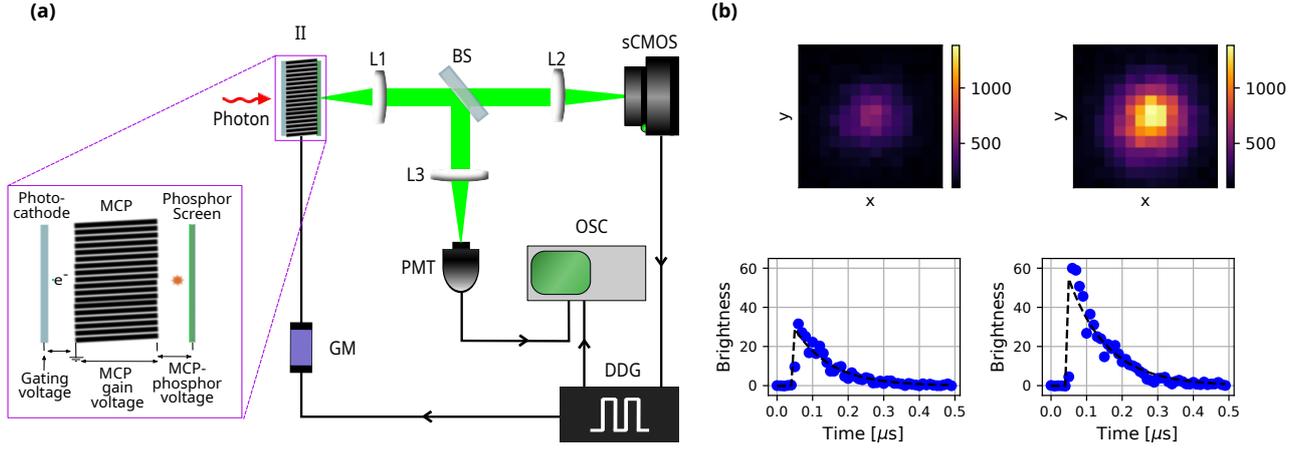}
%\begin{minipage}{\linewidth}
   \caption[0.8\linewidth]
   {
   Hybrid camera setup for measuring amplitudes correlations (\textbf{a}) and examples of signals registered by the detectors (\textbf{b}). \textbf{(a)} As a consequence of a single photon striking the image intensifier, a burst of photons (shown as the green beam) is emitted from the intensifier's output window. The beam is then split into two parts by a beam splitter: one part illuminates the fast detector (PMT) and the other is imaged onto the high spatial resolution detector (sCMOS camera). A precise delay generator is triggered by the full exposure of the sCMOS camera frame to open the image intensifier for a predetermined time. The signal from the photomultiplier tube is simultaneously captured using a digital oscilloscope. The inset provides a detailed view of the image intensifier's internal structure. II -- image intensifier; BS -- beam-splitter; L1, L2, L3 -- lenses; PMT -- photomultiplier tube; sCMOS -- sCMOS camera; GM -- gating module; OSC -- oscilloscope; DDG -- digital delay generator. \textbf{(b)} Examples of signals registered by the sCMOS (top row) and PMT (bottom row). With the image intensifier's approximate 20$\%$ quantum efficiency in photon-to-burst conversion, the two detectors can measure photon bursts from each single photon, allowing for the correlation of events. An example of the signals corresponding to a bright event are shown in the right column, a less bright event is depicted in the left column.}
%\end{minipage}
\label{fig:expt_amp_corr}
\end{figure*}

Single photon cameras are a powerful tool for measuring high-dimensional quantum correlations, as they can detect individual photons and capture images with high spatial and temporal resolution. These include electron multiplying charge coupled device (EMCCD)~\cite{Reichert2018, Burkhardt2006, Edgar2012, Daigle2009, Daigle2010}, and intensified scientific complementary metal oxide semiconductor (IsCMOS)~\cite{Chrapkiewicz:14, Dbrowski2018, Dbrowski2017} or intensified CCD cameras~\cite{Aspden2013, Fickler2013}, intensified high-speed Timepix cameras~\cite{Fisher_Levine_2016, Nomerotski_2017, Ianzano2020, https://doi.org/10.48550/arxiv.2302.03756} and single-photon avalanche diode (SPAD) arrays~\cite{Antolovic:18, Morimoto:20, Ndagano2020, Eckmann2020}. However, there is a trade-off between temporal and spatial resolution, as faster cameras tend to have lower spatial resolution and vice versa.

Intensified cameras offer high spatial resolution but suffer from a slow frame rate, limited by the CMOS or CCD sensor readout. To measure correlations of photons created at the same time, we need to gate the camera once per frame to access the temporal relation between detected events which results in a low-duty cycle. The gating time of the intensifier determines the coincidence window for photon detections, and the average photon flux should be kept significantly below one photon-pair per coincidence window to minimize the number of accidental coincidence detections. This results in a majority of frames being empty (no photon detections) requiring long measurement times for sufficient statistics. Therefore, the long measurement times and stability issues make the task of coincidence imaging such as the spatial correlation of photon pairs produced via SPDC challenging~\cite{Jachura:15, Chrapkiewicz:14}. 

Time-tagging SPAD arrays, on the other hand, have emerged as a promising alternative due to their higher temporal resolution, and continuous readout. Nonetheless, they typically have pixel counts smaller by orders of magnitude than cameras. For SPAD arrays, the sub-nanosecond temporal resolution is achieved through asynchronous readout, which limits the spatial resolution to tens, perhaps hundreds of pixels, and extending the resolution to the megapixel regime requires a gated operation~\cite{Antolovic:18, Morimoto:20, Ndagano2020, Eckmann2020}.

To overcome these limitations, we propose a novel hybrid intensified camera (HIC) that integrates a high spatial resolution sensor and a high temporal resolution detector. Our objective is to enable the detection and registration of individual photons, which is achieved by incorporating an image intensifier (II) capable of transforming a single photon detection event into a burst of photons while preserving its temporal and spatial information. With this HIC setup, we adaptively control the number of registered photons using feedback information from the fast (high temporal resolution) detector. Preliminary results of this work have been included in a patent~\cite{patent_cam} and presented at a conference~\cite{Kundu2022}.

We demonstrate the high-spatial-resolution and fast acquisition capabilities of the HIC setup by measuring non-classical correlations between photon pairs produced in a type-I SPDC process and comparing it to a standard intensifier camera.

\section{Single photon hybrid camera}
\subsection{Hybrid camera overview}
Our HIC utilizes off-the-shelf components and incorporates an image intensifier as in standard intensified cameras. The image intensifier is comprised of a photocathode, followed by a micro-channel plate (MCP), and a phosphor screen situated at the back of the intensifier (Fig.\ref{fig:expt_amp_corr}a - inset). When a photon illuminates the image intensifier, it can trigger the emission of a photoelectron from the photocathode. The photoelectron then enters one of nearly nine million channels of the MCP, accelerates due to the constant voltage across the plate, generates a macroscopic charge avalanche, and produces a bright flash of light at the phosphor screen.

The light from the image intensifier output is split into two detectors by a flat glass plate. A fraction of approximately 10$\%$ of the light is directed onto a photo-multiplier tube (PMT) that provides high temporal resolution, while the remaining 90$\%$ is imaged onto an sCMOS camera that offers a high spatial resolution.

We image the phosphor screen onto the sCMOS camera, eliminate the readout noise, and achieve sub-pixel spatial resolution by fitting each detected flash with a Gaussian function, recording its central position and amplitude~\cite{Jachura:15, Chrapkiewicz:14}. Remarkably, the spatial resolution of our hybrid camera surpasses that of the sCMOS camera component and is only constrained by the number of channels in the MCP of the image intensifier, which yields a spatial resolution of nearly nine megapixels.

\subsection{Amplitude correlations with two detectors}
%Standard intensified cameras suffer from low-duty cycles. By incorporating a second detector with a higher temporal resolution, the hybrid camera offers a solution to that. %low temporal resolution in IsCMOS cameras and allows for a more comprehensive image of the investigated phenomenon. 
The hybrid camera presents a solution to the low-duty cycle issue typically encountered in standard intensified cameras by incorporating a second detector with enhanced temporal resolution. Both the sCMOS camera and the high temporal resolution detector, PMT record the brightness of the phosphor screen flashes, which can be used to assign temporal information to the photons captured on the camera frame. In our approach (Fig.\ref{fig:expt_amp_corr}a), the light beam from the image intensifier (Hamamatsu V7090D-71-G272) is split on a beam splitter and simultaneously recorded by the sCMOS camera (Andor Zyla) and a photomultiplier tube (Hamamatsu R1924A). 

%\subsubsection{HIC operation}
When the camera initiates full-frame image acquisition, it sends a "Fire All" signal indicating that all rows of the sCMOS sensor are capturing data. This signal reaches the digital delay generator (Stanford Research DG645), which opens the image intensifier and triggers data recording from the photomultiplier on a digital oscilloscope (Teledyne LeCroy 610Zi) for a precisely determined time.
This setup yields frames from the sCMOS camera and corresponding voltage waveforms from the PMT shown in (Fig.\ref{fig:expt_amp_corr}b). By fitting appropriate functions to the amplitudes of both signals, we can extract the brightness of individual events, allowing us to correlate them in both the temporal and spatial domains. This, as we further demonstrate, enables us to perform time-tagging of individual photon detection events. Using this hybrid camera method, in section 4.3, as a proof-of-concept we show the potential of nanosecond time-tagging photon detections in a multi-megapixel spatial resolution sensor.
\begin{figure*}[ht!]
\centering
    \captionsetup{width=\linewidth}
    \centering
    \includegraphics[width=\linewidth]{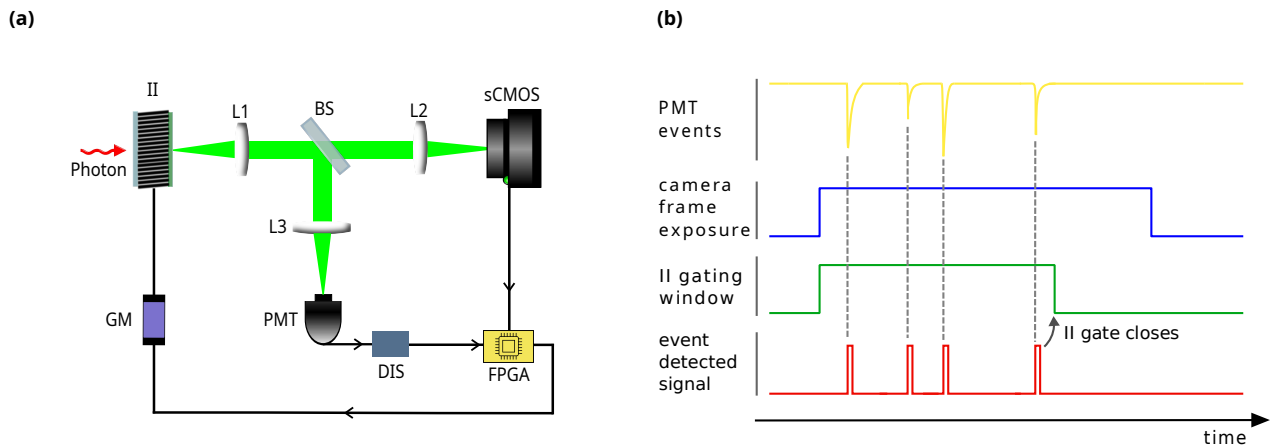}
 %\begin{minipage}{\linewidth}
   \caption[0.8\linewidth]{
   Representation of hybrid camera functionality: hardware (\textbf{a}) and software (\textbf{b}) illustration. \textbf{(a)} Schematic of the adaptive gating hybrid camera setup. After a photon strikes the image intensifier, it is converted to a burst of photons from the phosphor screen (shown as the green beam). The beam is then split into two parts by a beam splitter: one part illuminates the fast detector (PMT) and the other is imaged onto the high spatial resolution detector (sCMOS camera). Based on the information from the fast detector, FPGA can limit the gating time of the image intensifier, therefore controlling the number of registered photons. II -- image intensifier; BS -- beam-splitter; L1, L2, L3 -- lenses; PMT -- photomultiplier tube; DIS -- discriminator; sCMOS -- sCMOS camera; FPGA -- embedded device running the software; GM -- gating module. \textbf{(b)} Illustration of the adaptive time gating algorithm with counting events. This is an example of registering 4 pre-selected events: yellow trace represents the raw signal from the fast detector (PMT). The blue trace represents the sCMOS camera frame exposure while it registers the 4 events. Green trace represents the signal controlling the state of the image intensifier -- the high signal corresponds to the image intensifier being opened. The red trace represents the 4 events as counted by the FPGA. The image intensifier closed after registering 4 photons, therefore preventing any further registration of photons.}

%\end{minipage}
\label{fig:schem_adaptive_gating}
\end{figure*}

\subsection{Adaptively gated image intensifier}

By introducing a feedback loop for event detection, we can adaptively gate (i.e. control the exposure time of the image intensifier) and record a pre-selected number of photons on each frame (Fig.\ref{fig:schem_adaptive_gating}a).

To achieve this, the image intensifier opens simultaneously with the start of recording signals on both the sCMOS and PMT detectors (Fig.\ref{fig:schem_adaptive_gating}b). With the fast readout from the PMT, we can count the number of detected photons on the image intensifier much faster than the sCMOS camera's framerate. Since the image intensifier can be gated with nanosecond resolution, feedback from the PMT can disable the image intensifier after the detection of the set number of photons within nanoseconds (Fig.\ref{fig:schem_adaptive_gating}a--b). 

Since the gating and signal analysis introduce delays, the entire setup takes approximately 150 ns for the deactivation of the II after detecting the set number of photons. Custom electronics could reduce this delay to a few tens of nanoseconds. In order to register only events originating from the II, we use a fast analog voltage level discriminator. This allows us to reject noise and digitize the PMT signal. As a result, each detected photon is flagged with a TTL pulse at the output of the discriminator.

To enable fast electronic control of gating the image intensifier, we use an FPGA (NI myRIO-1900) that registers signals from both the discriminator and the camera acquisition. The FPGA registers a "Fire All" signal to ensure that all photons are registered when the camera is fully open -- all rows of the sCMOS camera are recording light. After detecting the "Fire All" signal, the FPGA opens the II and keeps the gate open until either a set number of photons is detected or the "Fire All" signal gets low, indicating that some rows of the camera stopped their acquisition. Ending the image intensifier's acquisition after registering a set number of photons prevents subsequent photons from being detected and allows us to register a preset number of photons. With our system, the gating time of the image intensifier is adjusted to match the rate of event detection in real-time, resulting in varying gate opening times at each frame, but the same number of photons detected.

\begin{figure*}[ht!]
\centering
    \captionsetup{width=\linewidth}
    \centering
    \includegraphics[width=\linewidth]{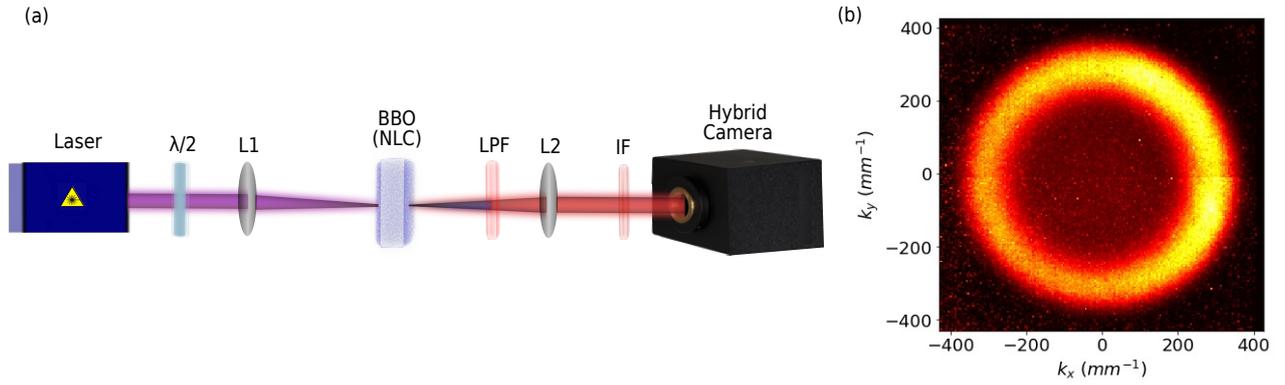}
 %\begin{minipage}{\linewidth}
   \caption[0.8\linewidth]{\textbf{(a)} Experimental setup for momentum correlation characterization of photon-pairs generated via type-I SPDC source. Pump laser (cw) -- 405 nm; BBO -- beta-barium borate (2 mm long); L1 and L2 - lenses of effective focal lengths of 400 mm and 120 mm respectively; LPF -- a combination of two long pass filters; IF -- Interference filter at $810 \pm 2$ nm. \textbf{(b)} The measured intensity distribution of the down-converted photons in the far field, where $k_{x}$ and $k_{y}$ are the detected transverse momenta respectively.
   }
%\end{minipage}
\label{fig:SPDC_expt_ring}
\end{figure*}

\section{Experimental setup for characterizing momentum correlations of down-converted photon pairs}

Here as an application of adaptive time gating of the hybrid camera setup, we investigate the far-field distribution of the SPDC light~\cite{Walborn2010} and observe the momentum anti-correlation of the generated bi-photon state in the adaptive gating mode of the HIC camera. The setup we use for our experiment is depicted in  Fig.\ref{fig:SPDC_expt_ring}a. Spatially entangled photon pairs are produced via a non-collinear, type-I spontaneous parametric down-conversion process in a 2 mm long beta-barium borate (BBO) nonlinear crystal pumped with a 405 nm laser beam of approximately 20 mW from a continuous wave diode laser. In order to meet the phase-matching condition of the SPDC process, a half-wave plate ($\lambda/2$) is used to set the polarization of the pump. The spatially filtered pump beam is focused onto the BBO crystal with the lens $L_1$ of focal length 400 mm. Both signal and idler photons generated as a pair by the pump beam of diameter 480 $\mu$m $(1/e^{2})$ on the crystal plane are of the same central wavelength (810 nm) and have the same polarization. After the crystal, a combination of two long-pass filters is used to block the pump beam. In the far-field configuration, these degenerate photons propagate through a composite Fourier lens system with an effective focal length of $f_e = 120$ mm which is represented by the lens $L_2$ in Fig.\ref{fig:SPDC_expt_ring}a. This composite system is made of three lenses of focal lengths 75 mm, 125 mm, and 200 mm respectively. The Fourier plane of the output of the BBO crystal is imaged onto the photocathode of the adaptively gated image intensifier. Before detection, the photon pairs %emitted around the degeneracy 
are spectrally selected by a narrow-band (810 $\pm$ 2 nm) interference filter (IF). 

\section{Experimental results}
\subsection{Photon transverse momentum correlation measurements}

\begin{figure*}[ht!]
\centering
    \captionsetup{width=\linewidth}
    \centering
    \includegraphics[width=\linewidth]{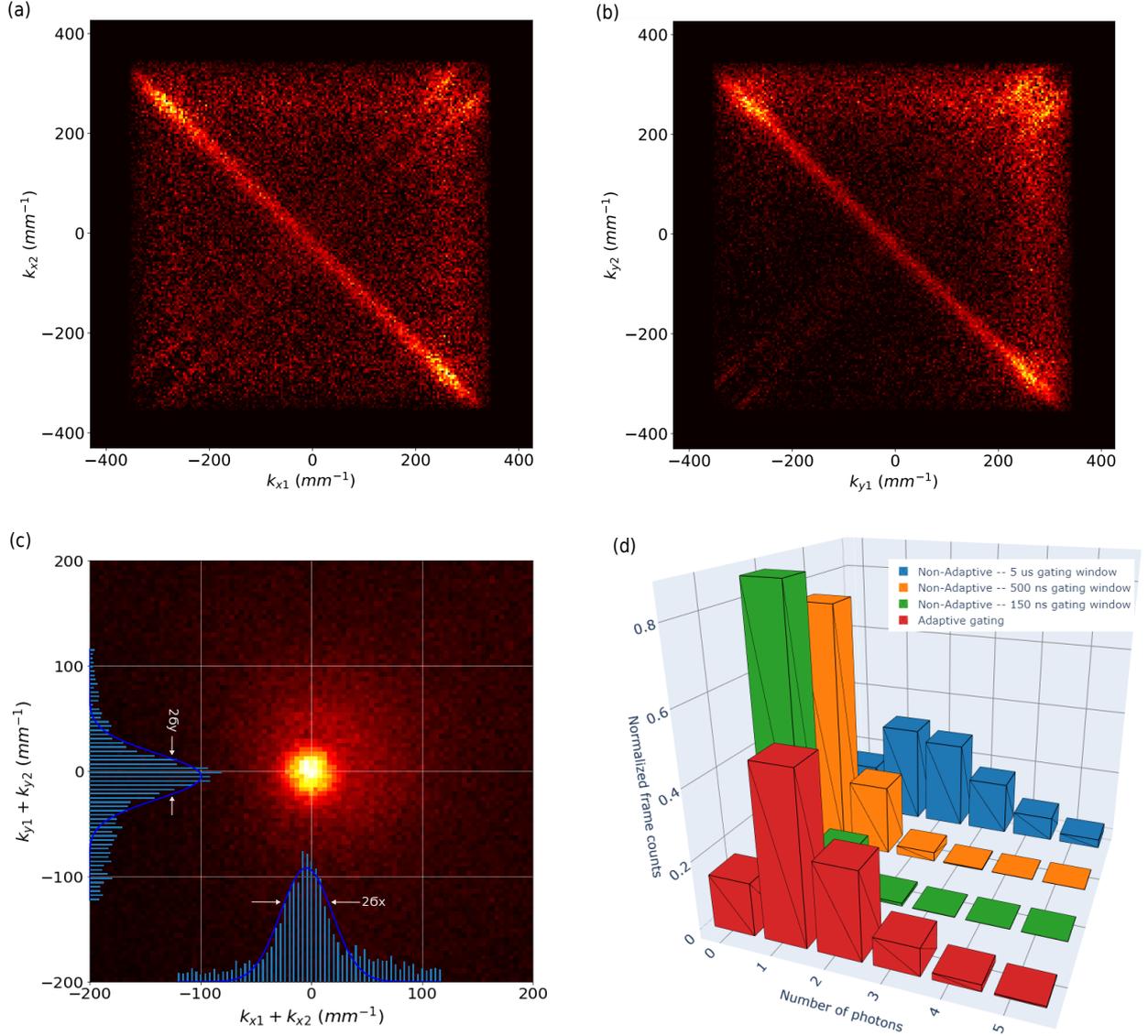}
 %\begin{minipage}{\linewidth}
   \caption[0.8\linewidth]{Results of photon transverse momentum correlation measurements. \textbf{(a-b)} Measured two-photon momentum correlations between signal and idler photons along $k_{x}$ and $k_{y}$ -- the subscripts 1, 2 represent signal and idler generated in SPDC. Correlation plots are represented after background correlation subtraction. \textbf{(c)} Measured correlation peak at the center of the transverse momentum sum-coordinates projection. The central peak is fitted with a Gaussian function to obtain its center and widths (standard deviations) along both $k_{x}$ and $k_{y}$, where $\sigma_{x}$ is $24.43$ mm$^{-1}$ and  $\sigma_{y}$ is $22.67$ mm$^{-1}$. \textbf{(d)} represents a comparison of frame numbers with specific photon counts in two scenarios - one with adaptive gating and one with non-adaptive gating. The experiment produced single-photon pairs through the SPDC which were registered with an identical setup. By triggering a single photon detection on a camera we should ideally observe two photons, as they come in pairs, and observe the corresponding bar to be the highest. The relative probabilities of registering one, two, and more photons were influenced by losses and the low quantum efficiency of the image intensifier. Considering one and two-photon frames as indicative of the adaptive gating success, we achieve a success rate of approximately 77.6$\%$ frames triggered correctly in adaptive gating, while in the case of non-adaptive gating with 150 ns gate window, the success rate is only around $13.7\%$.}
%\end{minipage}
\label{fig:KX_KY}
\end{figure*}

The experimental results of photon transverse momentum correlation measurements are shown in Fig.\ref{fig:KX_KY}a--c. We verified the correctness of gating by measuring the momentum correlations of signal and idler photons in the Fourier space -- the correlated photons have opposite transverse momenta and therefore lie on opposite sides of the center of the beam creating a ring. In Fig.\ref{fig:SPDC_expt_ring}b, we see the typical far-field SPDC type-I intensity pattern (ring). We can visualize it on two graphs, one per dimension (Fig.\ref{fig:KX_KY}a--b). These intense anti-diagonals represent the momentum anti-correlation of the photon pairs.

Because of the low quantum efficiency of the photocathode of the HIC and the presence of high background, we need to remove accidental coincidences to obtain a clear two-photon momentum anti-correlation. This is achieved by subtracting the correlation between photon pairs detected on consecutive frames from the measured correlation between photon pairs detected within the same frames. The peak of the correlation at the center of the sum coordinate projection (Fig.\ref{fig:KX_KY}c) is a signature of strong anti-correlations between two photons imposed by phase-matching (momentum conservation) in the pair production process via SPDC.

When an incident photon strikes the photocathode of the image intensifier it can release an electron. This electron is then accelerated by the gated potential towards the MCP. Upon impact with the MCP wall, secondary emission electrons are emitted initiating the avalanche gain process. Some electrons can be deflected backward from the MCP, re-entering different microchannels, and inducing additional electron avalanches. These events, known as cross-talk, can cause strong, artificial correlations as they are registered within the same camera frame. To remove such artificial correlations from our data, we discard photons detected too close (within 100 pixels of the camera) to each other. Implementing this cross-talk subtraction technique is vital when using an MCP for spatially resolved detection, as it helps to eliminate artificial correlations from our data~\cite{Lipka2018}. 

We obtained correlations for approximately 194 spatial modes on the momentum basis. To estimate the number of correlated modes, we divided the area of the intensity ring by the square of the correlation length. The correlation length corresponds to the size of the spatial mode in the far field, which we determined by fitting a Gaussian function to the correlation peak in the sum coordinate and extracting the standard deviations $\sigma_{x}$ and $\sigma_{y}$. %full width at half maximum (FWHM). 
This result indicates how many distinct spatial modes are in the intensity ring, which represents the spatial distribution of the down-converted photons~\cite{Schneeloch2016}.

\subsection{Comparison of adaptive vs. non-adaptive gating in photon counting}

Here, we provide a comparison of performance between the adaptively gated hybrid camera setup and the non-adaptive setup with an intensified sCMOS camera where the image intensifier is gated once per camera frame. The time between consecutive frames is typically $\sim$6 orders of magnitude longer than the minimum gating time of an IsCMOS camera~\cite{Jachura:15, Chrapkiewicz:14} (an IsCMOS camera can typically obtain a framerate of around 100 Hz at a full spatial resolution with a gate length in the nanoseconds range). For photon correlation measurements, we only register photons within the intensifier gating time, which determines the effective coincidence window for detected photons. 

In standard approaches (non-adaptive gating setup), we gate the image intensifier externally by sending signals from a digital pulse generator~\cite{Chrapkiewicz:14, Jachura:15}. In the adaptive gating approach the duration of the gating window is determined by the feedback loop. This technique allows for the observation of a single photon triggering the adaptive exposure and any additional photons registered within the coincidence window, which is the time between the first photon detection and the closing of the image intensifier (approximately 150 ns in our case). We compare adaptive gating with a standard (non-adaptive) gating with the gating time of 150 ns to achieve similar probabilities of registering false coincidences as in the adaptive case. Additionally, we conducted measurements using two other windows of the gating time (500 ns and 5 $\mu$s) as reference points when comparing our adaptive gating method to the non-adaptive approach with a 150 ns gating window.

The experimental results, presented in Fig.\ref{fig:KX_KY}d, compare the measured histograms of frame numbers with specific photon counts for the adaptive gating approach (purple bars) and standard non-adaptive gating cases (green, red, and blue bars for 150 ns, 500 ns, and 5 $\mu$s gating windows, respectively). In both cases, the objective is to observe single photon-pair creation events. In the case of non-adaptive gating, we observed that the majority of recorded frames were empty (see green bars), which was necessary to maintain low signal levels and prevent accidental coincidence detections, mostly originating from separate photon-pair generation events. The comparison of experimental outcomes for the number of detected photons per frame in both adaptive and non-adaptive methods (Fig.\ref{fig:KX_KY}d) demonstrates the potential of the adaptive gating approach to enhance data acquisition rates.

As our focus is on observing single photon-pair creation events, we set our camera for single photon detection. The image intensifier gate closes after detecting the first photon, as, in SPDC, photons are generated simultaneously. Assuming no losses in the setup and a 100$\%$ quantum efficiency of the detector, we would anticipate most counts to be pairs from the SPDC, with only a small number of multi-photon events consisting of additional noise photons or multiple pair detections. However, due to factors such as electrical noise, dark counts, the PMT detects some false events that are not registered by the sCMOS camera, resulting in some empty frames in the histogram for adaptive gating (Fig.\ref{fig:KX_KY}d). Furthermore, the low (below 20$\%$) quantum efficiency of the image intensifier means that coincidences are not always observed; hence, we consider one and two-photon frames as indicative of the adaptive gating success. Consequently, we achieved a success rate of approximately 77.6$\%$ frames triggered correctly, while the non-adaptive gating (with a 150 ns gating window) scenario yielded a success rate of only $13.7\%$; an improvement of around 5.6 times. Adaptive gating reduces the occurrence of empty frames by almost 5.4 times compared to the non-adaptive gating scenario. Moreover, increasing the gating window size in the non-adaptive approach results in an increased false coincidence rate despite the lower number of empty frames present (see Fig.\ref{fig:KX_KY}d). 
\begin{figure*}[ht!]
\centering
    \captionsetup{width=\linewidth}
    \centering
    \includegraphics[width=\linewidth]{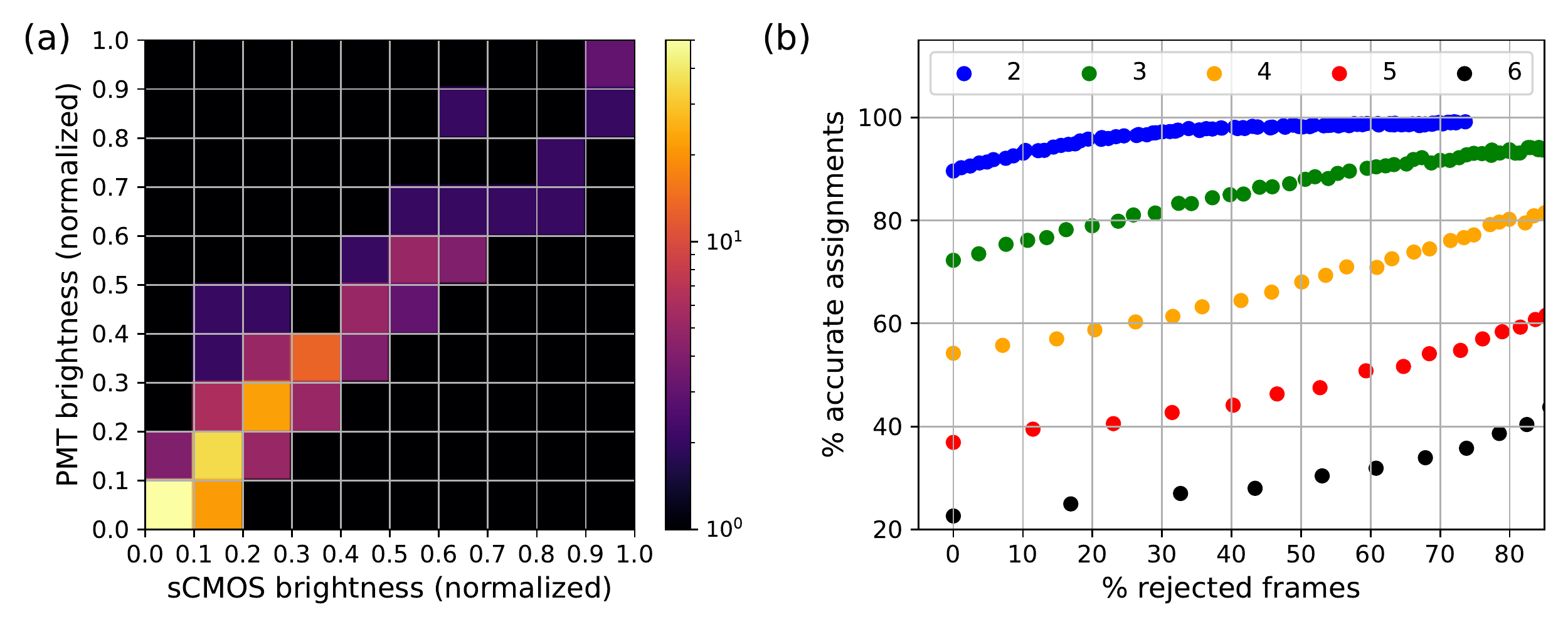}
 %\begin{minipage}{\linewidth}
   \caption[0.8\linewidth]{Time tagging capabilities of the hybrid camera based on brightness correlation. (a): sCMOS and PMT brightness correlation map of recorded single photon frames, plotted in logarithmic scale. Color scale represents the number of occurrences, with black indicating at most 1 occurrence. (b): Single photon time tagging accuracy as a function of percentage of rejected frames for two (blue), three (green), four (orange), five (red) and six (black) photon frames.}
%\end{minipage}
\label{fig:corrfig}
\end{figure*}
\subsection{Time-tagging single photons}
Correlating the brightness of events recorded by both HIC detectors enables determining the arrival time of each photon registered on the sCMOS camera frame. It can easily be combined with adaptive gating, which allows for the acquisition of frames with a predetermined number of accurately timestamped photons.
To verify the time tagging capabilities and determine its accuracy in multi-photon frames, a synthetic data set was constructed from camera frames containing a single photon. For each of the single photon frames, two numbers corresponding to PMT and sCMOS brightness were recorded. Their distribution and linear correlation are shown in Fig \ref{fig:corrfig}a. Next, n-photon frames were constructed by drawing n-element tuples without replacement from the set of single photon frames. When the phosphor flashes are too close in brightness, the pair (or tuple) can be rejected to maintain high time tagging accuracy. Each tuple could contain either correct or incorrect time tagging, or be neglected due to the brightness of the events being too close in value. The $0\%$ rejected frames in Fig. \ref{fig:corrfig}c corresponds to not discarding any events based on the brightness criterion, maintaining $90\%$ time tagging accuracy for frames containing 2 photons. Tightening the brightness bound, $80\%$ accuracy can be obtained for 3-photon frames while neglecting $20\%$ of frames, without significant impact on the measurement time. Majority of the events exhibit relatively low brightness. A more uniform brightness distribution can be achieved by appropriately sweeping the image intensifier voltage, reducing the rejected frames percentage.

\section{Conclusions and outlook}
In summary, we demonstrated a single-photon sensitive hybrid camera that allows a selection of a number of detected photons by adaptively controlling the image intensifier. The main advantage of our approach is a significant increase in data acquisition speed with respect to standard non-adaptive approaches. We showed the possibility of the gate-time adjustment to the rate of photon detection through a fast feedback loop based on FPGA.
This active approach surpasses the statistical one which can only predict the expected value of a parameter, in our case an opening time of the image intensifier. Our approach breaks this limitation by continuously counting photons in real-time and using this count to terminate the ongoing acquisition process. It benefits from the dual measurement nature of the device and allows us to fine-tune the acquisition length immediately after photons detection by the fast detector.

We have shown that it is possible to utilize our detection technique to observe quantum optical phenomena, for instance, the SPDC process, more effectively. 
As an additional application of the HIC we can use it to time-tag each detected spatially resolved photon by the sCMOS camera. We can achieve this by comparing the amplitudes of the photons detected by both PMT and sCMOS. This capability enhances the functionality of our hybrid camera, making it an even more versatile tool for various imaging applications, bridging the gap between the temporal resolution of time tagging SPAD arrays and the spatial resolution of intensified cameras~\cite{patent_cam}. 

Our hybrid camera overcomes the challenges faced by conventional intensified cameras by adding a fast feedback loop and offering a multi-megapixel resolution unobtainable by the time-tagging SPAD arrays at the moment. It can be used in fundamental quantum information science and its applications such as quantum imaging~\cite{Defienne2021, Ndagano2020, Ndagano2022, DAngelo2004}, quantum walks~\cite{Peruzzo2010, Defienne2016}, Boson sampling~\cite{2019} and studies of quantum correlations in scattering media~\cite{Peeters2010, Lib2022}.

\begin{backmatter}
\bmsection{Acknowledgments and Funding} 
We acknowledge discussions with Rados\l{}aw
Chrapkiewicz, Wojciech Wasilewski, Konstantin Rusakov, Anat Daniel, Sanjay Kapoor, and Ali Golestani. This work was supported by the Foundation for Polish Science under the FIRST TEAM project 'Spatiotemporal photon correlation measurements for quantum metrology and super-resolution microscopy' co-financed by the European Union under the European Regional Development Fund (POIR.04.04.00-00-3004/17-00).

\bmsection{Disclosures} RL, JS, GF, AKP are inventors on EP3648453~\cite{patent_cam}.

\bmsection{Data Availability Statement} The data and the codes used to process the data are available from the corresponding author upon reasonable request.

\end{backmatter}

\bibliography{sample}

\begin{thebibliography}{10}
\newcommand{\enquote}[1]{``#1''}

\bibitem{Walborn2006}
S.~P. Walborn, D.~S. Lemelle, M.~P. Almeida, and P.~H.~S. Ribeiro,
  \enquote{Quantum key distribution with higher-order alphabets using spatially
  encoded qudits,} {\protect\JournalTitle{Physical Review Letters}} \textbf{96}
  (2006).

\bibitem{PhysRevA.77.062323}
S.~P. Walborn, D.~S. Lemelle, D.~S. Tasca, and P.~H. Souto~Ribeiro,
  \enquote{Schemes for quantum key distribution with higher-order alphabets
  using single-photon fractional fourier optics,} {\protect\JournalTitle{Phys.
  Rev. A}} \textbf{77}, 062323 (2008).

\bibitem{Walborn2007}
S.~P. Walborn, D.~S. Ether, R.~L. de~Matos~Filho, and N.~Zagury,
  \enquote{Quantum teleportation of the angular spectrum of a single-photon
  field,} {\protect\JournalTitle{Physical Review A}} \textbf{76} (2007).

\bibitem{PhysRevLett.123.070505}
Y.-H. Luo, H.-S. Zhong, M.~Erhard, X.-L. Wang, L.-C. Peng, M.~Krenn, X.~Jiang,
  L.~Li, N.-L. Liu, C.-Y. Lu, A.~Zeilinger, and J.-W. Pan, \enquote{Quantum
  teleportation in high dimensions,} {\protect\JournalTitle{Phys. Rev. Lett.}}
  \textbf{123}, 070505 (2019).

\bibitem{PhysRevX.3.011013}
G.~A. Howland and J.~C. Howell, \enquote{Efficient high-dimensional
  entanglement imaging with a compressive-sensing double-pixel camera,}
  {\protect\JournalTitle{Phys. Rev. X}} \textbf{3}, 011013 (2013).

\bibitem{Farooq2022}
A.~Farooq, U.~Khalid, J.~ur~Rehman, and H.~Shin, \enquote{Robust quantum state
  tomography method for quantum sensing,} {\protect\JournalTitle{Sensors}}
  \textbf{22}, 2669 (2022).

\bibitem{PhysRevA.83.052325}
D.~S. Tasca, R.~M. Gomes, F.~Toscano, P.~H. Souto~Ribeiro, and S.~P. Walborn,
  \enquote{Continuous-variable quantum computation with spatial degrees of
  freedom of photons,} {\protect\JournalTitle{Phys. Rev. A}} \textbf{83},
  052325 (2011).

\bibitem{BechmannPasquinucci2000}
H.~Bechmann-Pasquinucci and W.~Tittel, \enquote{Quantum cryptography using
  larger alphabets,} {\protect\JournalTitle{Physical Review A}} \textbf{61}
  (2000).

\bibitem{Ecker2019}
S.~Ecker, F.~Bouchard, L.~Bulla, F.~Brandt, O.~Kohout, F.~Steinlechner,
  R.~Fickler, M.~Malik, Y.~Guryanova, R.~Ursin, and M.~Huber,
  \enquote{Overcoming noise in entanglement distribution,}
  {\protect\JournalTitle{Physical Review X}} \textbf{9} (2019).

\bibitem{Reimer2018}
C.~Reimer, S.~Sciara, P.~Roztocki, M.~Islam, L.~Romero~Cortés, Y.~Zhang,
  B.~Fischer, S.~Loranger, R.~Kashyap, A.~Cino, S.~T. Chu, B.~E. Little, D.~J.
  Moss, L.~Caspani, W.~J. Munro, J.~Azaña, M.~Kues, and R.~Morandotti,
  \enquote{High-dimensional one-way quantum processing implemented on d-level
  cluster states,} {\protect\JournalTitle{Nature Physics}} \textbf{15},
  148--153 (2018).

\bibitem{Erhard2020}
M.~Erhard, M.~Krenn, and A.~Zeilinger, \enquote{Advances in high-dimensional
  quantum entanglement,} {\protect\JournalTitle{Nature Reviews Physics}}
  \textbf{2}, 365--381 (2020).

\bibitem{Malik2016}
M.~Malik, M.~Erhard, M.~Huber, M.~Krenn, R.~Fickler, and A.~Zeilinger,
  \enquote{Multi-photon entanglement in high dimensions,}
  {\protect\JournalTitle{Nature Photonics}} \textbf{10}, 248--252 (2016).

\bibitem{Krenn2014}
M.~Krenn, M.~Huber, R.~Fickler, R.~Lapkiewicz, S.~Ramelow, and A.~Zeilinger,
  \enquote{Generation and confirmation of a (100 {\texttimes} 100)-dimensional
  entangled quantum system,} {\protect\JournalTitle{Proceedings of the National
  Academy of Sciences}} \textbf{111}, 6243--6247 (2014).

\bibitem{Defienne2021}
H.~Defienne, B.~Ndagano, A.~Lyons, and D.~Faccio, \enquote{Polarization
  entanglement-enabled quantum holography,} {\protect\JournalTitle{Nature
  Physics}} \textbf{17}, 591--597 (2021).

\bibitem{Moreau2019}
P.-A. Moreau, E.~Toninelli, T.~Gregory, and M.~J. Padgett, \enquote{Imaging
  with quantum states of light,} {\protect\JournalTitle{Nature Reviews
  Physics}} \textbf{1}, 367--380 (2019).

\bibitem{Tse2019}
M.~T. et~al., \enquote{Quantum-enhanced advanced {LIGO} detectors in the era of
  gravitational-wave astronomy,} {\protect\JournalTitle{Physical Review
  Letters}} \textbf{123} (2019).

\bibitem{PhysRevLett.104.060401}
T.~V\'ertesi, S.~Pironio, and N.~Brunner, \enquote{Closing the detection
  loophole in bell experiments using qudits,} {\protect\JournalTitle{Phys. Rev.
  Lett.}} \textbf{104}, 060401 (2010).

\bibitem{Reichert2018}
M.~Reichert, H.~Defienne, and J.~W. Fleischer, \enquote{Massively parallel
  coincidence counting of high-dimensional entangled states,}
  {\protect\JournalTitle{Scientific Reports}} \textbf{8} (2018).

\bibitem{Burkhardt2006}
M.~Burkhardt and P.~Schwille, \enquote{Electron multiplying {CCD} based
  detection for spatially resolved fluorescence correlation spectroscopy,}
  {\protect\JournalTitle{Optics Express}} \textbf{14}, 5013 (2006).

\bibitem{Edgar2012}
M.~Edgar, D.~Tasca, F.~Izdebski, R.~Warburton, J.~Leach, M.~Agnew, G.~Buller,
  R.~Boyd, and M.~Padgett, \enquote{Imaging high-dimensional spatial
  entanglement with a camera,} {\protect\JournalTitle{Nature Communications}}
  \textbf{3} (2012).

\bibitem{Daigle2009}
O.~Daigle, C.~Carignan, J.-L. Gach, C.~Guillaume, S.~Lessard, C.-A. Fortin, and
  S.~Blais-Ouellette, \enquote{Extreme faint flux imaging with an {EMCCD},}
  {\protect\JournalTitle{Publications of the Astronomical Society of the
  Pacific}} \textbf{121}, 866--884 (2009).

\bibitem{Daigle2010}
O.~Daigle and S.~Blais-Ouellette, \enquote{Photon counting with an {EMCCD},} in
  \emph{{SPIE} Proceedings,}  E.~Bodegom and V.~Nguyen, eds. ({SPIE}, 2010).

\bibitem{Chrapkiewicz:14}
R.~Chrapkiewicz, W.~Wasilewski, and K.~Banaszek, \enquote{High-fidelity
  spatially resolved multiphoton counting for quantum imaging applications,}
  {\protect\JournalTitle{Opt. Lett.}} \textbf{39}, 5090--5093 (2014).

\bibitem{Dbrowski2018}
M.~D{\k{a}}browski, M.~Mazelanik, M.~Parniak, A.~Leszczy{\'{n}}ski, M.~Lipka,
  and W.~Wasilewski, \enquote{Certification of high-dimensional entanglement
  and einstein-podolsky-rosen steering with cold atomic quantum memory,}
  {\protect\JournalTitle{Physical Review A}} \textbf{98} (2018).

\bibitem{Dbrowski2017}
M.~D{\k{a}}browski, M.~Parniak, and W.~Wasilewski,
  \enquote{Einstein{\textendash}podolsky{\textendash}rosen paradox in a hybrid
  bipartite system,} {\protect\JournalTitle{Optica}} \textbf{4}, 272 (2017).

\bibitem{Aspden2013}
R.~S. Aspden, D.~S. Tasca, R.~W. Boyd, and M.~J. Padgett, \enquote{{EPR}-based
  ghost imaging using a single-photon-sensitive camera,}
  {\protect\JournalTitle{New Journal of Physics}} \textbf{15}, 073032 (2013).

\bibitem{Fickler2013}
R.~Fickler, M.~Krenn, R.~Lapkiewicz, S.~Ramelow, and A.~Zeilinger,
  \enquote{Real-time imaging of quantum entanglement,}
  {\protect\JournalTitle{Scientific Reports}} \textbf{3} (2013).

\bibitem{Fisher_Levine_2016}
M.~Fisher-Levine and A.~Nomerotski, \enquote{{TimepixCam}: a fast optical
  imager with time-stamping,} {\protect\JournalTitle{Journal of
  Instrumentation}} \textbf{11}, C03016--C03016 (2016).

\bibitem{Nomerotski_2017}
A.~Nomerotski, I.~Chakaberia, M.~Fisher-Levine, Z.~Janoska, P.~Takacs, and
  T.~Tsang, \enquote{Characterization of {TimepixCam}, a fast imager for the
  time-stamping of optical photons,} {\protect\JournalTitle{Journal of
  Instrumentation}} \textbf{12}, C01017--C01017 (2017).

\bibitem{Ianzano2020}
C.~Ianzano, P.~Svihra, M.~Flament, A.~Hardy, G.~Cui, A.~Nomerotski, and
  E.~Figueroa, \enquote{Fast camera spatial characterization of photonic
  polarization entanglement,} {\protect\JournalTitle{Scientific Reports}}
  \textbf{10} (2020).

\bibitem{https://doi.org/10.48550/arxiv.2302.03756}
B.~Courme, C.~Vernière, P.~Svihra, S.~Gigan, A.~Nomerotski, and H.~Defienne,
  \enquote{Quantifying high-dimensional spatial entanglement with a
  single-photon-sensitive time-stamping camera,} ArXiv:2302.03756 [quant-ph]
  (2023).

\bibitem{Antolovic:18}
I.~M. Antolovic, C.~Bruschini, and E.~Charbon, \enquote{Dynamic range extension
  for photon counting arrays,} {\protect\JournalTitle{Opt. Express}}
  \textbf{26}, 22234--22248 (2018).

\bibitem{Morimoto:20}
K.~Morimoto, A.~Ardelean, M.-L. Wu, A.~C. Ulku, I.~M. Antolovic, C.~Bruschini,
  and E.~Charbon, \enquote{Megapixel time-gated spad image sensor for 2d and 3d
  imaging applications,} {\protect\JournalTitle{Optica}} \textbf{7}, 346--354
  (2020).

\bibitem{Ndagano2020}
B.~Ndagano, H.~Defienne, A.~Lyons, I.~Starshynov, F.~Villa, S.~Tisa, and
  D.~Faccio, \enquote{Imaging and certifying high-dimensional entanglement with
  a single-photon avalanche diode camera,} {\protect\JournalTitle{npj Quantum
  Information}} \textbf{6} (2020).

\bibitem{Eckmann2020}
B.~Eckmann, B.~Bessire, M.~Untern\"{a}hrer, L.~Gasparini, M.~Perenzoni, and
  A.~Stefanov, \enquote{Characterization of space-momentum entangled photons
  with a time resolving {CMOS} {SPAD} array,} {\protect\JournalTitle{Optics
  Express}} \textbf{28}, 31553 (2020).

\bibitem{Jachura:15}
M.~Jachura and R.~Chrapkiewicz, \enquote{Shot-by-shot imaging of
  hong--ou--mandel interference with an intensified scmos camera,}
  {\protect\JournalTitle{Opt. Lett.}} \textbf{40}, 1540--1543 (2015).

\bibitem{patent_cam}
{Radoslaw Lapkiewicz, Jerzy Szuniewicz, Grzegorz Firlik, Wojciech Wasilewski,
  Wojciech Zwolinski, Piotr Wegrzyn, and Aleksander Krupinski-Ptaszek},
  \enquote{System for detection with high temporal and spatial resolution and
  method for detection with high temporal and spatial resolution,} in
  \emph{European Patent Office, EP3648453,}  (May, 2020).

\bibitem{Kundu2022}
S.~Kundu, J.~Szuniewicz, G.~Firlik, and R.~Lapkiewicz, \enquote{Single-photon
  camera with an adaptively gated image intensifier,} in \emph{Frontiers in
  Optics + Laser Science 2022 ({FIO}, {LS}),}  (Optica Publishing Group, 2022).

\bibitem{Walborn2010}
S.~Walborn, C.~Monken, S.~P{\'{a}}dua, and P.~S. Ribeiro, \enquote{Spatial
  correlations in parametric down-conversion,} {\protect\JournalTitle{Physics
  Reports}} \textbf{495}, 87--139 (2010).

\bibitem{Lipka2018}
M.~Lipka, M.~Parniak, and W.~Wasilewski, \enquote{Microchannel plate cross-talk
  mitigation for spatial autocorrelation measurements,}
  {\protect\JournalTitle{Applied Physics Letters}} \textbf{112}, 211105 (2018).

\bibitem{Schneeloch2016}
J.~Schneeloch and J.~C. Howell, \enquote{Introduction to the transverse spatial
  correlations in spontaneous parametric down-conversion through the biphoton
  birth zone,} {\protect\JournalTitle{Journal of Optics}} \textbf{18}, 053501
  (2016).

\bibitem{Ndagano2022}
B.~Ndagano, H.~Defienne, D.~Branford, Y.~D. Shah, A.~Lyons, N.~Westerberg,
  E.~M. Gauger, and D.~Faccio, \enquote{Quantum microscopy based on
  hong{\textendash}ou{\textendash}mandel interference,}
  {\protect\JournalTitle{Nature Photonics}} \textbf{16}, 384--389 (2022).

\bibitem{DAngelo2004}
M.~D'Angelo, Y.-H. Kim, S.~P. Kulik, and Y.~Shih, \enquote{Identifying
  entanglement using quantum ghost interference and imaging,}
  {\protect\JournalTitle{Physical Review Letters}} \textbf{92} (2004).

\bibitem{Peruzzo2010}
A.~Peruzzo, M.~Lobino, J.~C.~F. Matthews, N.~Matsuda, A.~Politi, K.~Poulios,
  X.-Q. Zhou, Y.~Lahini, N.~Ismail, K.~W\"{o}rhoff, Y.~Bromberg, Y.~Silberberg,
  M.~G. Thompson, and J.~L. OBrien, \enquote{Quantum walks of correlated
  photons,} {\protect\JournalTitle{Science}} \textbf{329}, 1500--1503 (2010).

\bibitem{Defienne2016}
H.~Defienne, M.~Barbieri, I.~A. Walmsley, B.~J. Smith, and S.~Gigan,
  \enquote{Two-photon quantum walk in a multimode fiber,}
  {\protect\JournalTitle{Science Advances}} \textbf{2} (2016).

\bibitem{2019}
\enquote{Photonic implementation of boson sampling: a review,}
  {\protect\JournalTitle{Advanced Photonics}} \textbf{1}, 1 (2019).

\bibitem{Peeters2010}
W.~H. Peeters, J.~J.~D. Moerman, and M.~P. van Exter, \enquote{Observation of
  two-photon speckle patterns,} {\protect\JournalTitle{Physical Review
  Letters}} \textbf{104} (2010).

\bibitem{Lib2022}
O.~Lib and Y.~Bromberg, \enquote{Quantum light in complex media and its
  applications,} {\protect\JournalTitle{Nature Physics}} \textbf{18}, 986--993
  (2022).

\end{thebibliography}

\end{multicols}
\end{document}